# Tuning the Competing magnetic interactions in RTiGe (R = Tb, Er) Compounds and tailoring the Magnetocaloric effect

R. Nirmala

*Tata Institute of Fundamental Research, Mumbai 400-005, India*

S. K. Malik[*]

*International Centre for Condensed Matter Physics - ICCMP, University of Brasilia, Brasilia, Brazil*

**Abstract**

The tetragonal layered compounds TbTiGe and ErTiGe order antiferromagnetically at 276 K and 39 K, respectively. Partial substitution of Mo for Ti in these two compounds modifies the magnetic interactions giving rise to a ferromagnetic ground state which results in an enhanced magnetocaloric effect. The magnetic entropy change in $ErTi_{0.85}Mo_{0.15}Ge$ for a magnetic field change of 5 T is ~10.5 J/kg/K against ~0.8 J/kg/K for ErTiGe in the vicinity of the magnetic transition. Thus, magnetocaloric properties of such layered materials may be tunable by suitable chemical substitutions.

[*] Corresponding author.
skm@tifr.res.in

Recently, the equiatomic RTiGe (R = Rare earth) compounds (tetragonal CeFeSi-type) have received considerable attention because of their high Néel temperatures which range from 70 K for R = Pr to 412 K for R = Gd [1, 2]. The transition temperatures of the heavier rare earth compounds of this series follow the so-called de Gennes scaling. Neutron diffraction studies on TbTiGe and ErTiGe have shown that these compounds have collinear antiferromagnetic structure, with ferromagnetic R-layers coupled antiferromagnetically along the c-axis [3]. The large span of magnetic ordering temperatures in RTiGe compounds makes them as good candidates for the study of magnetocaloric effect (MCE) [4-6]. Field induced magnetic transitions have been observed in DyTiGe and HoTiGe and a magnetoresistance of ~ 20% is reported for the former, near its metamagnetic transition (MMT) [7].

The structure of RTiGe compounds is characterized by the occurence of R-Ge-Ti-Ti-Ge-R layers propagating along the tetragonal c-axis where intra-plane and inter-plane R-R spacings are expected to play a vital role in determining the dominant exchange interaction and hence the magnetic ordering in these compounds. In addition, magnetoresistance measurements on intermetallic compounds of layered structure have been sought, in order to elucidate the mechanism of giant magnetoresistance in artificial magnetic multilayers, etc. In this context, we present and discuss here the results of magnetization, magnetoresistance and heat capacity studies on TbTiGe and ErTiGe. These compounds show a moderate MCE in spite of their layered structure and the presence of competing magnetic interactions. However, partial substitution of Ti by Mo, modifies the magnetic interactions in favour of ferromagnetism and leads to more than 10-fold increase in MCE.

Polycrystalline samples of TbTiGe, ErTiGe, $TbTi_{0.85}Mo_{0.15}Ge$ and $ErTi_{0.85}Mo_{0.15}Ge$ were prepared by electric arc melting under argon atmosphere starting from stoichiometric amounts of high purity constituent elements (Tb, Er – 99.9 % pure, Ti, Mo – 99.99 % pure and Ge - 99.999 % pure, Cerac Inc. USA). The samples were remelted several times to ensure their homogeneity and further annealed at 1000°C for 5 days in vacuum. The samples were characterized by powder X-ray diffraction experiments at room temperature. Magnetization measurements on these samples were carried out in the temperature range of 2 – 300 K using commercial magnetometers (MPMS and PPMS, Quantum Design, USA). Heat capacity was measured by a relaxation technique in the same temperature range, in fields up to 5 T (PPMS).

Room temperature X-ray diffraction data confirm that the TbTiGe and ErTiGe are single phase compounds forming in the tetragonal, CeFeSi-type structure (space group *P4/nmm*, No. 129). Phase composition was confirmed by EDAX analysis. The lattice parameters of these compounds are given in Table 1. The rare earth, the transition metal and the p-element all have distinct lattice sites in this structure.

Magnetization (M) measurements on these compounds reveal that TbTiGe and ErTiGe order antiferromagnetically at 276 K and 39 K, respectively (Fig. 1 and 2). A second low temperature transition is observed in both the compounds at ~ 55 K and 10 K, respectively. The paramagnetic effective moment, $\mu_{eff}$, values correspond to the respective free rare earth ion moments. The observed paramagnetic Curie temperature, $\theta_P$, values are quite large and positive (253 K for TbTiGe and 40 K for ErTiGe). Since $\theta_P$ reflects the collective exchange interactions in the compound, its positive value implies that the ferromagnetic interactions are also present in these compounds. This observation is supported by recent neutron diffraction studies on RTiGe compounds [3] where it has been identified that the magnetic moments are indeed coupled ferromagnetically within the unit cell but the moments in adjacent unit cells are coupled antiferromagnetically along the c axis. The low temperature transition in both the compounds might be associated with a possible change from a collinear to a non-collinear antiferromagnetic structure and needs to be further investigated by low temperature neutron diffraction measurements at intermediate temperatures.

The virgin magnetization vs. field isotherm of TbTiGe obtained at 2 K is more or less linear in field as expected for an antiferromagnet (inset in Fig. 1). However, a large hysteresis and a field induced step in magnetization are observed in the envelope curve pointing out the role of coexisting ferromagnetic interactions. The magnetization reaches only a value of about 2 $\mu_B$/Tb$^{3+}$ in 7 T field. The M-H data of ErTiGe at 2 K show that this compound undergoes a metamagnetic transition at a critical field of ~ 1.5 T giving rise to a field-induced ferromagnetic state [inset in Fig. 2]. The saturation magnetization value is found to be 8.3 $\mu_B$/Er$^{3+}$ which is slightly lower than the free ion value of 9 $\mu_B$.

The heat capacity of TbTiGe shows a weak anomaly near $T_N$ {Fig. 3) which disappears on the application of a magnetic field of 5 T without involving appreciable entropy change near the transition. The anomaly in heat capacity is rather weak because it occurs at relatively higher temperatures where the phonon contribution to heat capacity is dominant. However, the heat capacity of ErTiGe exhibits a distinct peak at $T_N$ {Fig. 4). This peak shifts to lower temperatures with application of magnetic field as expected for antiferromagnets and vanishes in an applied field of 5 T. A weak anomaly is also seen at the lower transition temperature in this compound. Magnetocaloric effect has been calculated in terms of the isothermal magnetic entropy change, $\Delta S_m$, and the adiabatic temperature change, $\Delta T_{ad}$, using the following expression:

$$\Delta S_m(T,H) = \int_0^T \frac{C(T,H) - C(T,0)}{T} dT$$

where C(T, 0) and C(T, H) are the heat capacities of the sample measured at temperature T in zero field and in an applied field H. For a field change of 5 T, the value of $\Delta S_m$ near the magnetic transition is ~ 0.09 J/kg/K for TbTiGe and 0.75 J/kg/K for ErTiGe and the corresponding values of $\Delta T_{ad}$ are 1.2 K and 4 K, respectively [Figs. 5-6]. The low value of magnetic entropy change arises from the dominant antiferromagnetic interactions in these compounds. Also from this study, it is evident that the field induced, magnetic-only transition need not necessarily lead to a large variation in magnetic entropy across the transition, as compared to that of field-induced magnetic transition that is also associated with a structural transition in intermetallic compounds and alloys [8]. The $\Delta T_{ad}$ value for TbTiGe is quite small (1.2K for 5T field change) compared to that of elemental Tb (8 K for 5 T field change). The $\Delta T_{ad}$ value of elemental Er is 4 K for 6 T field change [9] and it has nearly the same value in ErTiGe for a field change of 5 T. The Er-based compound shows better magnetocaloric effect than the Tb-based compound because of the field induced antiferromagnetic to ferromagnetic transition that nearly aligns the spins along the field direction whereas antiferromagnetic interactions remains dominant in TbTiGe.

Partial substitution of Mo for Ti in RTiGe compounds gives rise to a tetragonal CeScSi-type crystal structure (space group *I4/mmm,* no. 139) for $TbTi_{0.85}Mo_{0.15}Ge$ and $ErTi_{0.85}Mo_{0.15}Ge$ [10]. The CeScSi-type structure is similar to the parent CeFeSi-type structure but for the difference in the alternating double layers of rare earth and p-element along the c-axis [Fig. 7]. The number of formula units per unit cell doubles leading to a near doubling of the c-lattice parameter [Table 1]. The Mo-substitution at the Ti-site leads to subtle changes in the R-R distances between and within the layers. The R-R interlayer and intralayer distances decrease as a result of this substitution. For example, in TbTiGe, the intralayer and interlayer Tb-Tb distances are 0.4050 nm and 0.3756 nm, respectively which become 0.4012 nm and 0.3668 nm after Mo-substitution. The ratio, (V/R), of the cell volume, V, to the corresponding rare earth radius, R, is smaller for $RTi_{0.85}Mo_{0.15}Ge$ compounds compared to that of the parent RTiGe compounds. Thus it is envisaged that the changes in R-R distances along the c-axis could lead to a change in the magnetic properties. This is borne out experimentally (vide-infra).

Magnetization measurements on $TbTi_{0.85}Mo_{0.15}Ge$, in an applied field 0.5 T indicate a ferromagnetic-like transition at ~308 K ($T_C$) (Fig. 8). Development of antiferromagnetic interactions is observed below 78 K where the magnetization decreases a little. The magnetization vs. magnetic field isotherm obtained at 2 K shows sharp metamagnetic steps that yield a magnetization of only ~4.6 $\mu_B/Tb^{3+}$ in applied fields of 7 T (inset in Fig. 8). On the other hand, the low field magnetization data on ErTi$_{0.85}$Mo$_{0.15}$Ge reveal a ferromagnetic transition at ~ 60 K (T$_C$) (Fig. 9). A small peak in low field magnetization is also seen for this compound around 10 K. The field dependence of magnetization is nearly ferromagnetic but for a weak S-shaped metamagnetic behaviour in the reverse magnetization curves (quadrant 2 and 4) (Fig. 9) with a magnetization value of ~8.2 μ$_B$/Er$^{3+}$ in 9 T field. Thus Mo-substitution at Ti-site gives rise to an enhancement of the prevailing ferromagnetic interactions in RTiGe compounds. This observation could qualitatively be understood in terms of change in interlayer interactions in the substituted compounds. Since the parent compounds are A-type antiferromagnets with ferromagnetic intralayer interactions that are antiferromagnetically coupled along c-axis, Mo-substitution has probably enhanced the overall ferromagnetic exchange interactions in the compound thus leading to an increased magnetic ordering temperature values. This notion is further supported by the recent neutron diffraction data on TbTi$_{0.85}$Mo$_{0.15}$Ge which reveal a collinear ferromagnetic and non-collinear ferromagnetic ordering of Tb-moments at the two inflection points in magnetization data [11].

To compare the MCE values of Mo-substituted compound with those of the parent member, we have measured several M-H isotherms in the temperature range of 25 K to 80 K for ErTi$_{0.85}$Mo$_{0.15}$Ge (Fig. 10). The magnetic entropy change has been calculated by numerically integrating the Maxwell relation, $(\partial S / \partial H)_T = (\partial M / \partial T)_H$,

$$\Delta S_M (T,H) = S_M (T,H) - S_M (T,0) = \int_0^H \left( \frac{\partial M}{\partial T} \right)_H dH \qquad (1).$$

Making a numerical approximation to Eq. 1, ΔS$_M$, can be calculated from isothermal magnetization measurements performed at small, discrete field intervals at different temperatures near T$_C$, using the formula,

$$|\Delta S_M| = \sum_i \frac{M_i - M_{i+1}}{T_{i+1} - T_i} \Delta H \qquad (2)$$

where $M_i$ and $M_{i+1}$ are the values of magnetization at temperatures $T_i$ and $T_{i+1}$, respectively.

The variation of magnetic entropy ($\Delta S_M$) with temperature for $ErTi_{0.85}Mo_{0.15}Ge$ in the temperature range of 25 K to 80 K, for various applied fields, is shown in Fig. 11. A maximum magnetic entropy change of ~10.5 J/Kg/K near $T_C$ is observed for this compound which is more than an order of magnitude higher than that in the parent ErTiGe. This observation is in concordance with the assumption that ferromagnetic interactions are stronger in the Mo-substituted compound. Thus in a compound with interplay of positive and negative exchange interactions, a chemical substitution leading to a substantial change in the unit cell dimensions is being exploited to tune the effective magnetic interactions and hence to improve the magnetocaloric effect.

In summary, partial substitution of Mo for Ti in TbTiGe and ErTiGe enhances the ferromagnetic interactions in these compounds. The magnetocaloric effect near the magnetic transition is enhanced by an order of magnitude by this substitution. It is possible that similar substitutions in known compounds of layered structure with coexisting magnetic interactions and giant MCE may lead to further improvement of the magnetocaloric property, by proper tuning of the desired dominant ferromagnetic exchange.

**Acknowledgements**

The authors thank S. A. Watpade for his help in arc-melting the samples, B.A. Chalke for EDAX analysis and A. V. Morozkin, Moscow Lomonosov State University for useful discussions on the crystal structure.

Table 1

Lattice parameters, a and c, antiferromagnetic ordering temperature, $T_N$, of RTiGe (R = Tb, Er) and ferromagnetic ordering temperature ($T_C$) of Mo-substituted compounds

| Sample | a (nm) | c (nm) | $T_N/T_C$ (K) |
|---|---|---|---|
| TbTiGe | 0.4059 | 0.7667 | 276 |
| ErTiGe | 0.4019 | 0.7548 | 39 |
| TbTi$_{0.85}$Mo$_{0.15}$Ge | 0.4015 | 1.5316 | 308 |
| ErTi$_{0.85}$Mo$_{0.15}$Ge | 0.3992 | 1.5105 | 60 |

**Figure Captions**

Fig. 1. Magnetization vs. Temperature of TbTiGe compound [Inset: Magnetization vs. Field isotherm at 2 K in fields up to 7 T]

Fig. 2 Magnetization vs. Temperature of ErTiGe compound [Inset: Magnetization vs. Field isotherm at 2 K in fields up to 7 T]

Fig. 3. Heat capacity vs. Temperature of TbTiGe compound as a function of applied field

Fig. 4. Heat capacity vs. Temperature of ErTiGe compound as a function of applied field

Fig. 5. Adiabatic temperature change of TbTiGe compound near the magnetic transition, for an applied field change.

Fig. 6. Adiabatic temperature change of ErTiGe compound near the magnetic transition, for an applied field change.

Fig. 7. The sequence of double-layer Tb atoms alternation along c axis in the (a) CeScSi-type structure (TbTi$_{0.85}$Mo$_{0.15}$Ge) and (b) CeFeSi-type structure (TbTiGe) [D1 – Intralayer R-R distance, D2 – interlayer R-R distance]

Fig. 8. Magnetization vs. Temperature of TbTi$_{0.85}$Mo$_{0.15}$Ge compound [Inset: Magnetization vs. Field isotherm at 2 K in fields up to 7 T or 9 T]

Fig. 9 Magnetization vs. Temperature of ErTi$_{0.85}$Mo$_{0.15}$Ge compound [Inset: Magnetization vs. Field isotherm at 2 K in fields up to 7 T or 9 T]

Fig. 10. Magnetization vs. field isotherms obtained in the temperature range of 25 K – 80 K for ErTi$_{0.85}$Mo$_{0.15}$Ge compound

Fig. 11. Magnetic entropy change near the ferromagnetic transition in ErTi$_{0.85}$Mo$_{0.15}$Ge compound for various applied fields

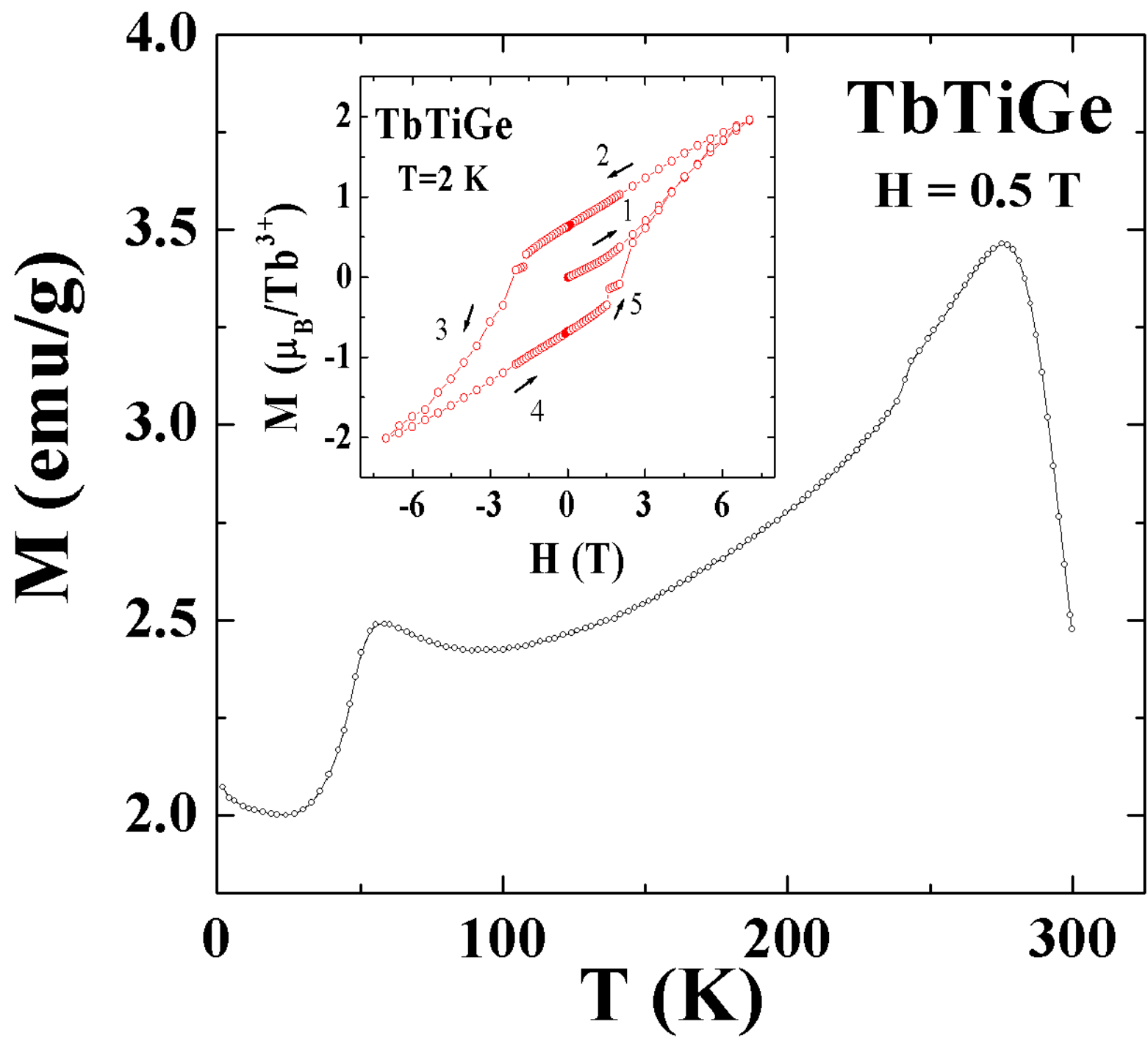

Fig. 1

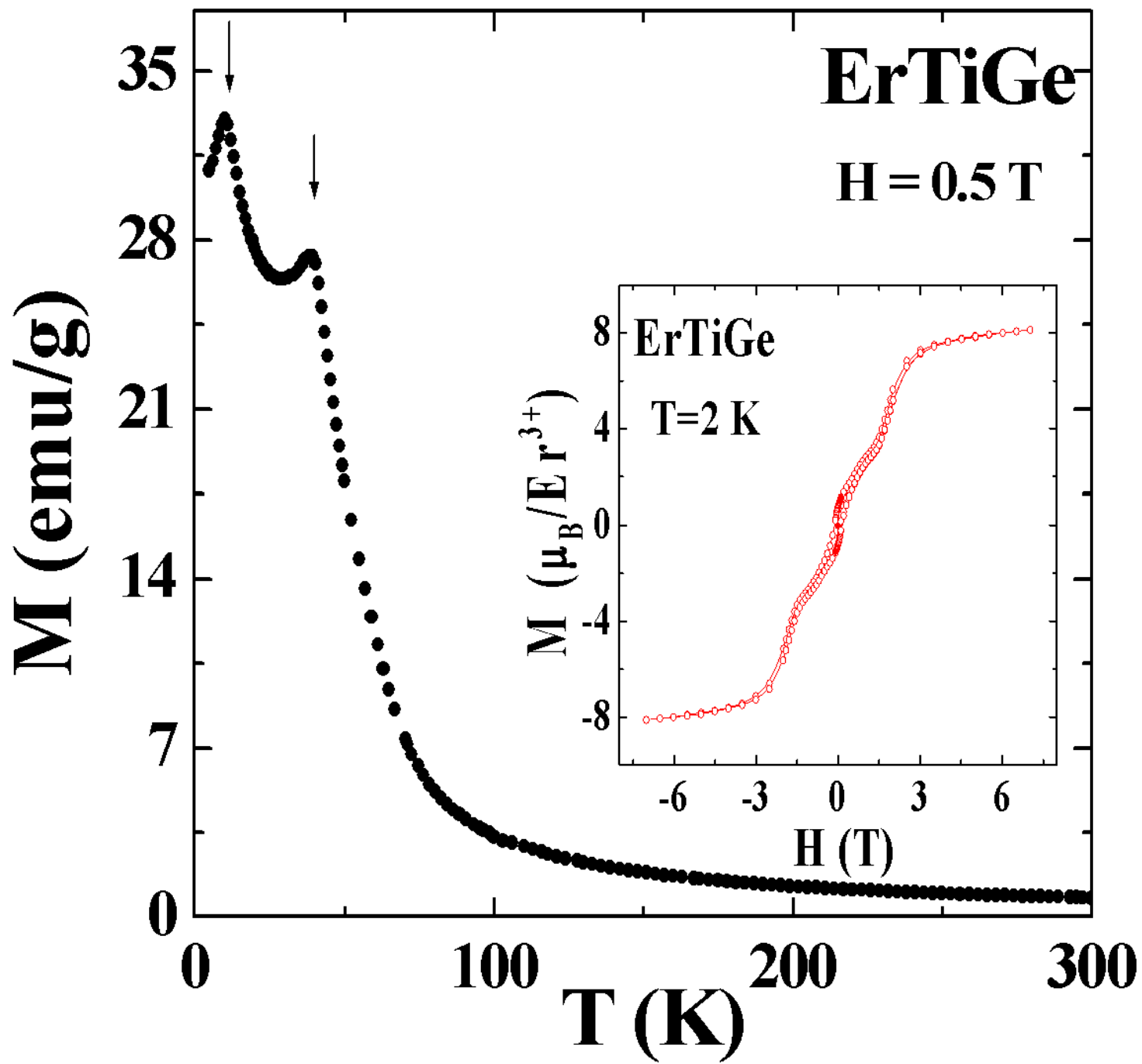

Fig. 2

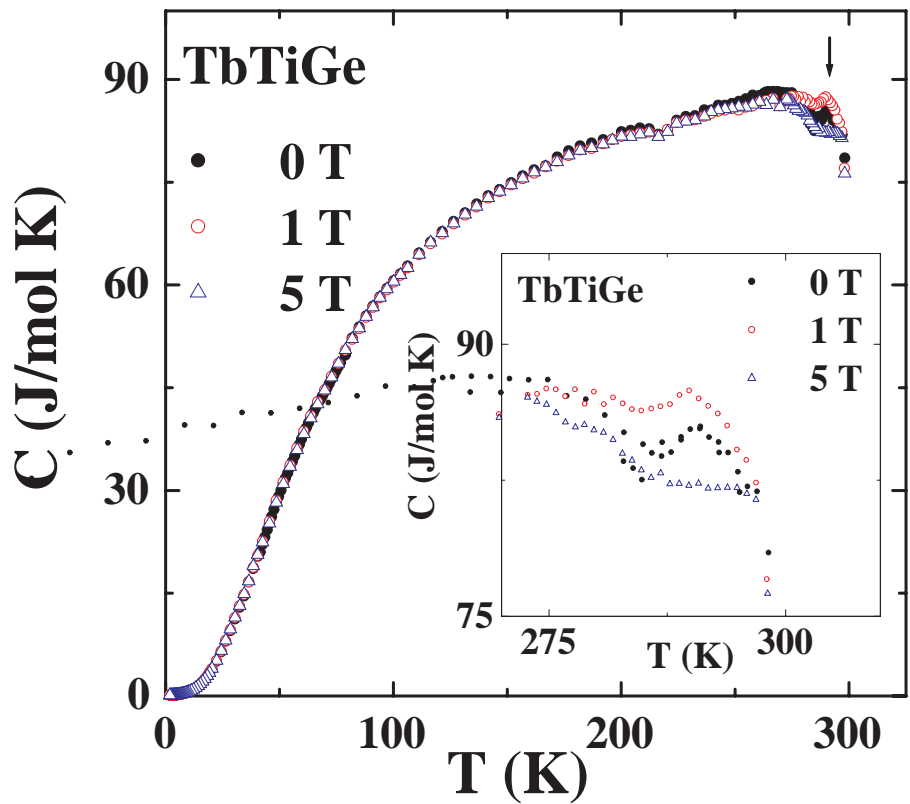

Fig. 3

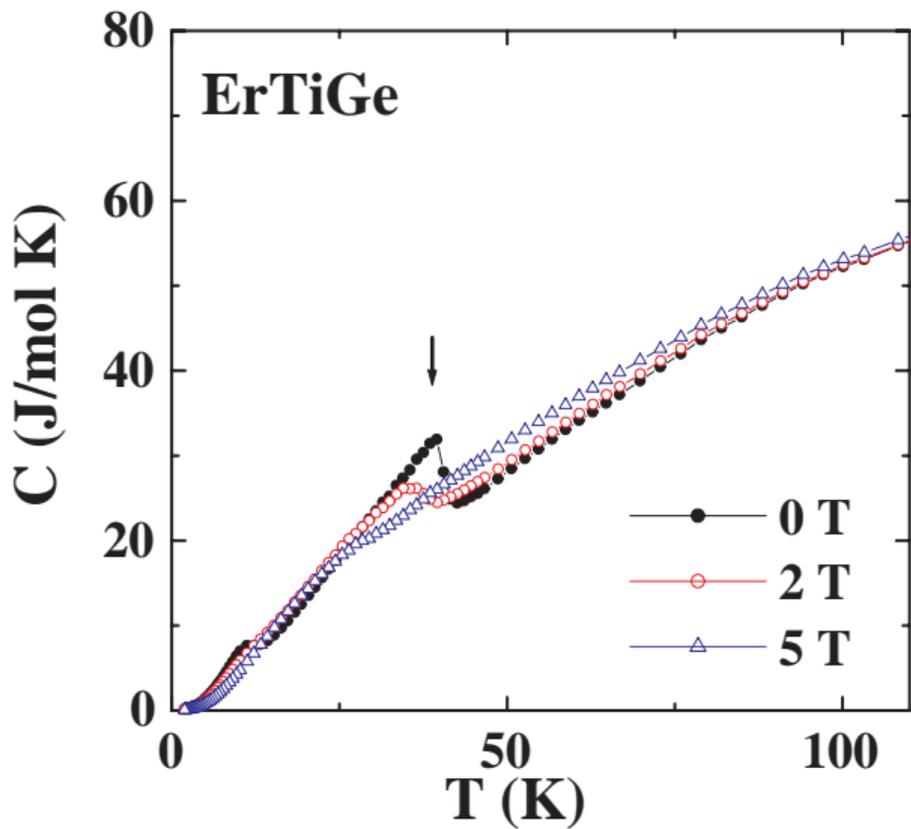

Fig. 4

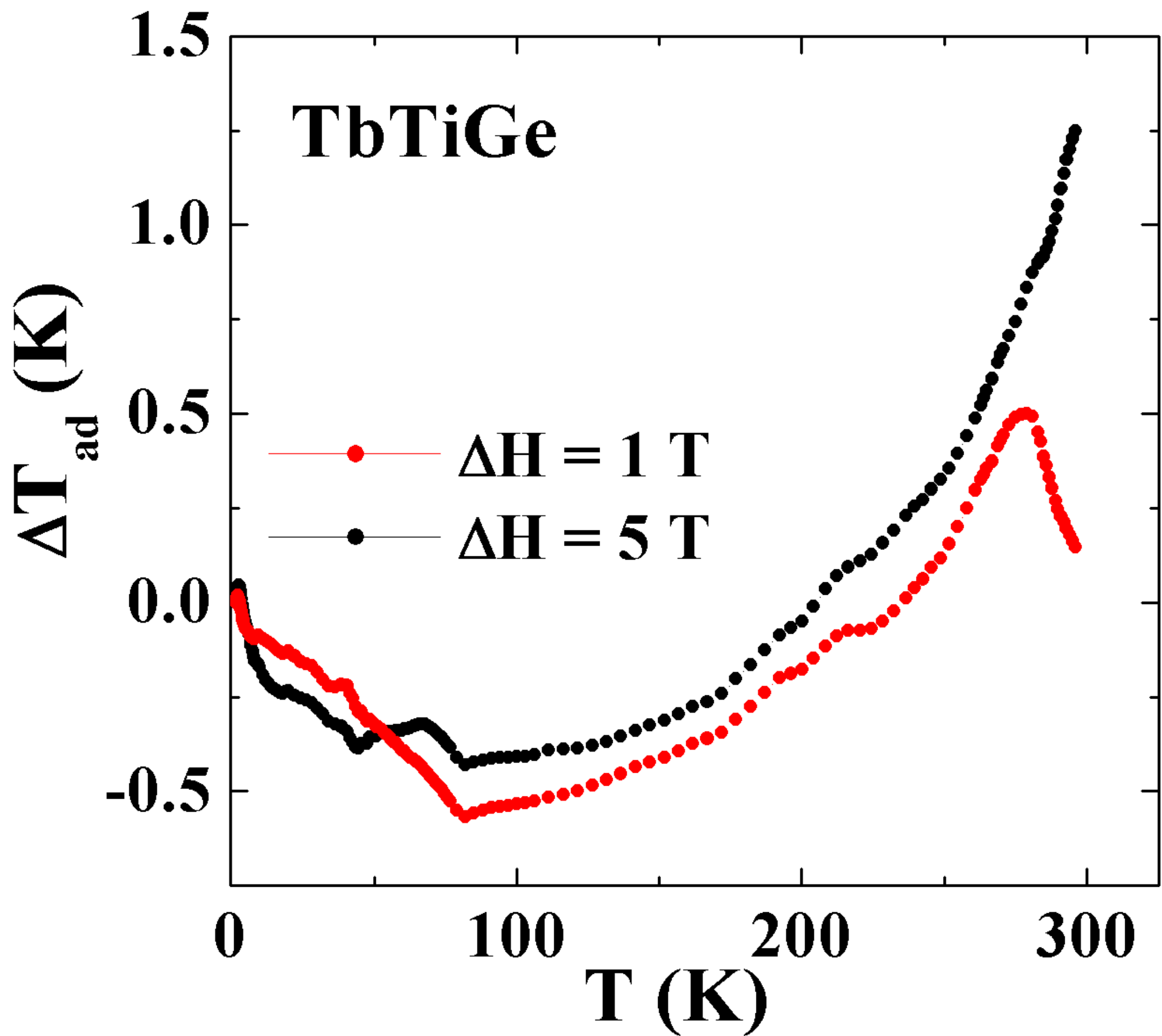

Fig. 5

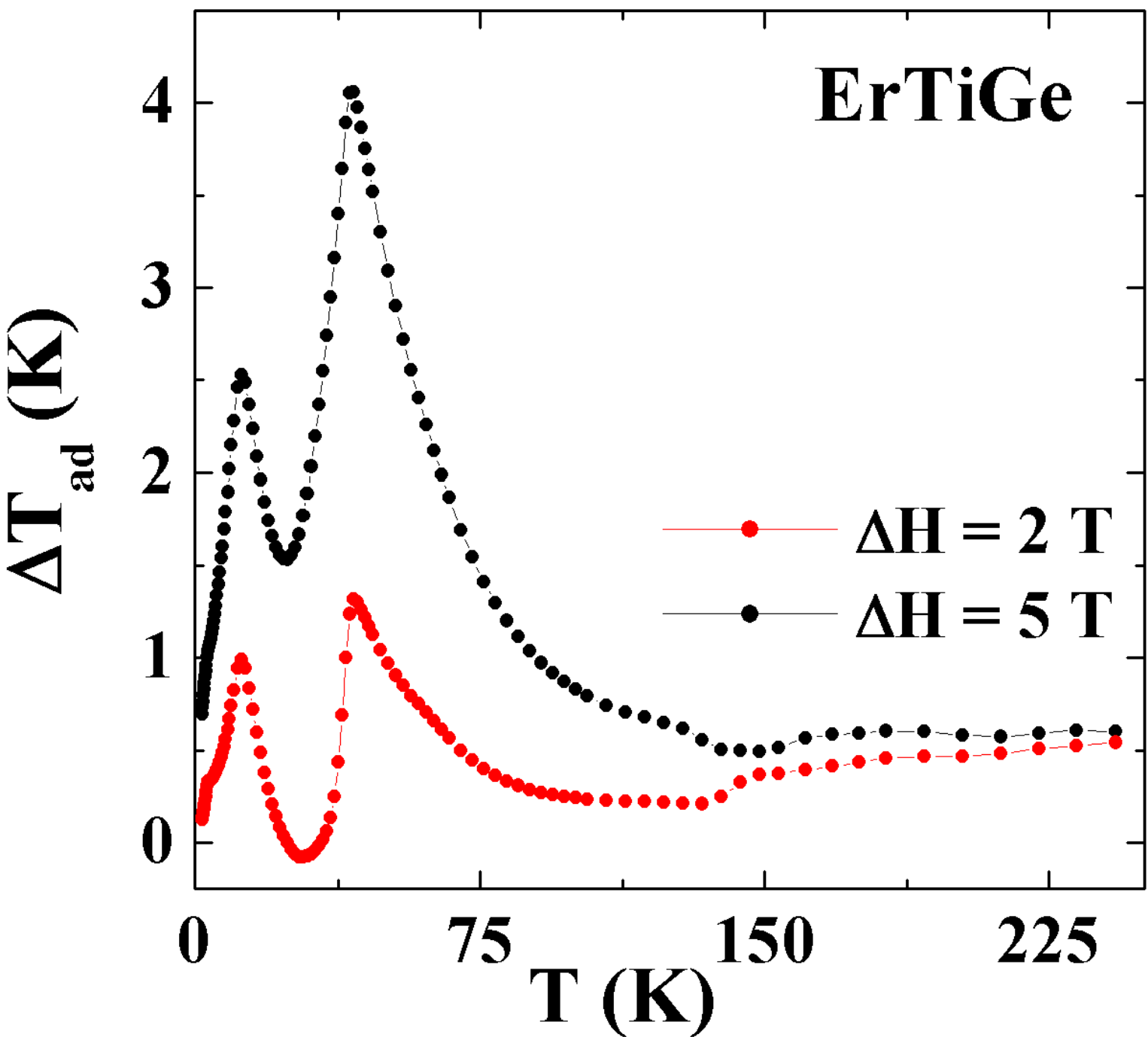

Fig. 6

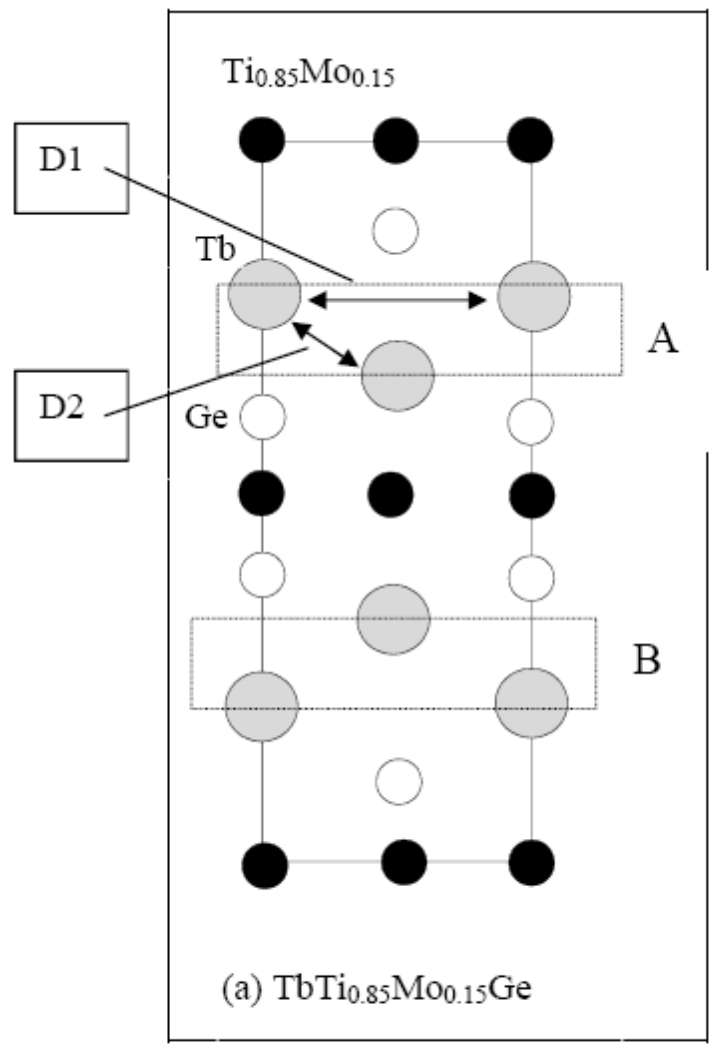 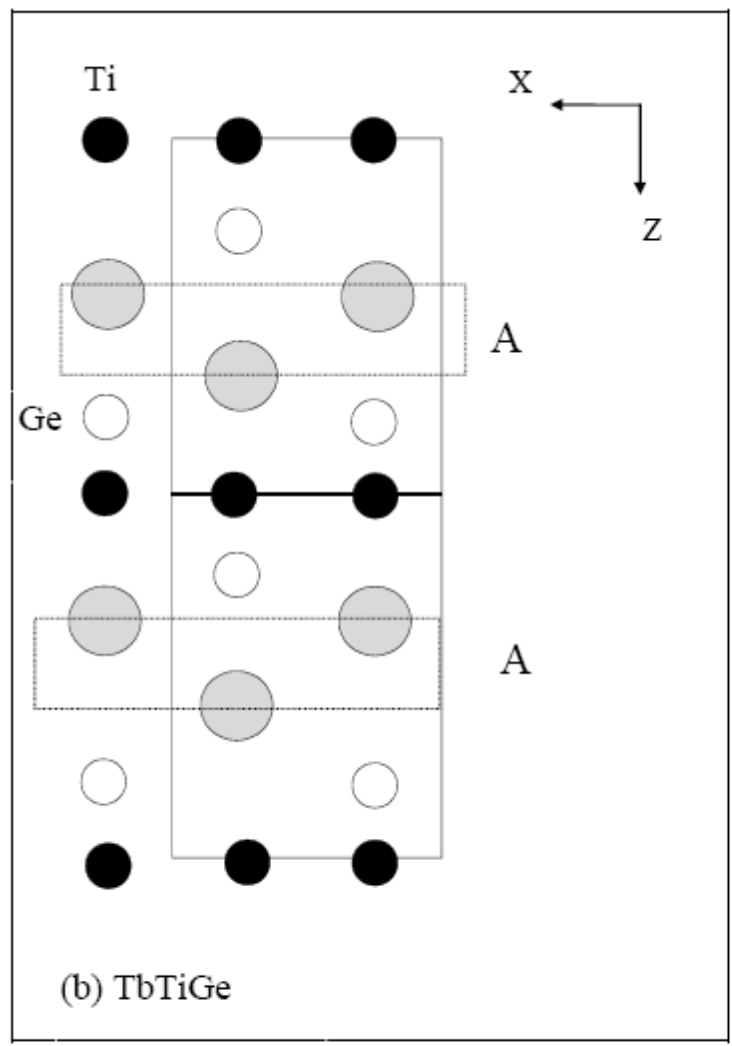

Fig. 7

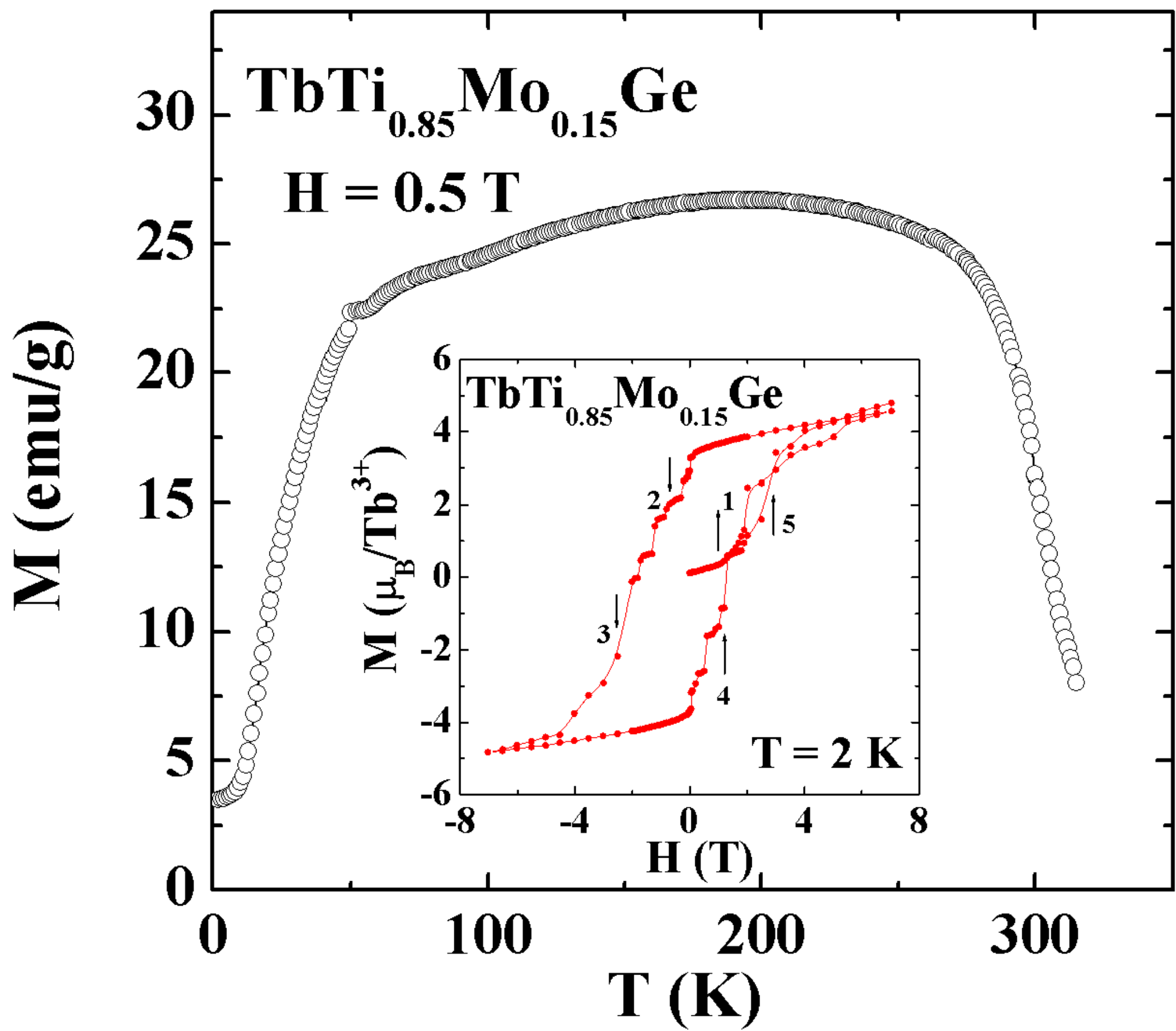

Fig. 8

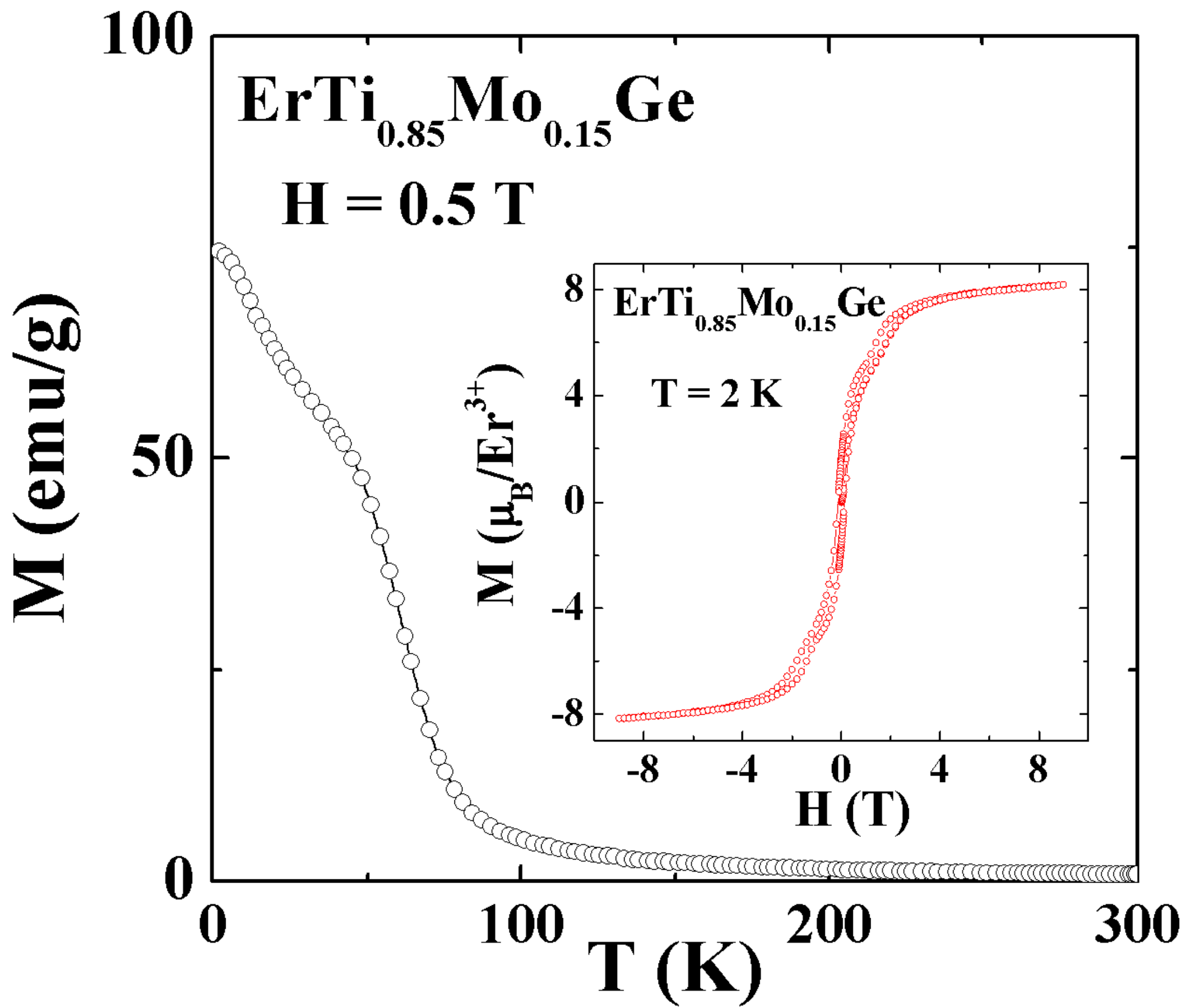

Fig. 9

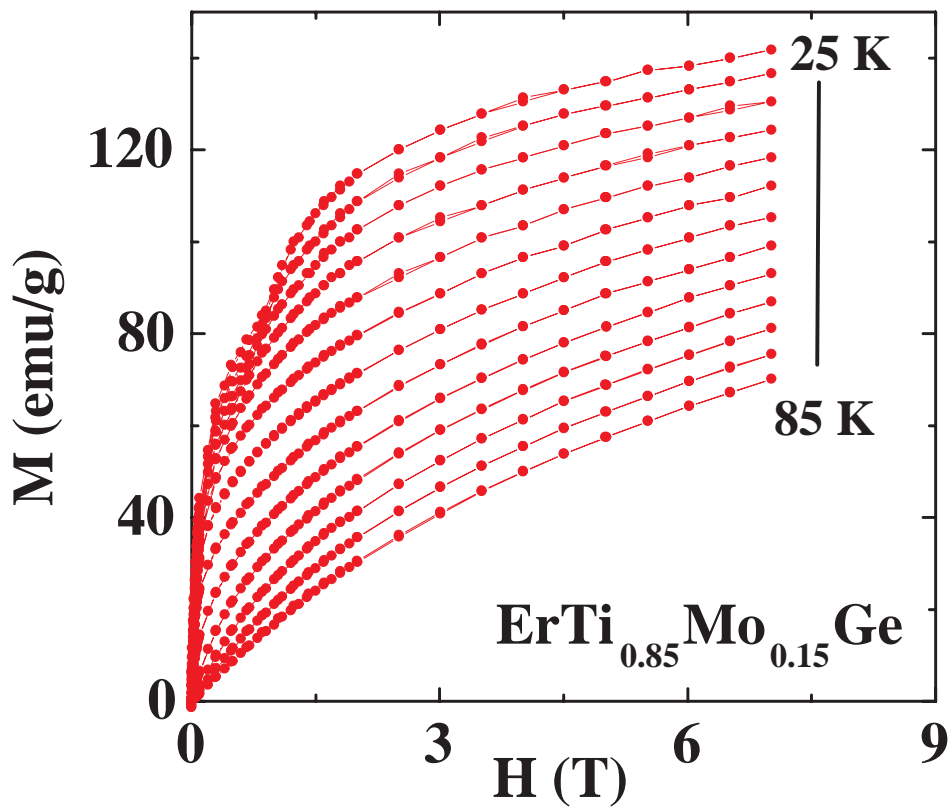

Fig. 10

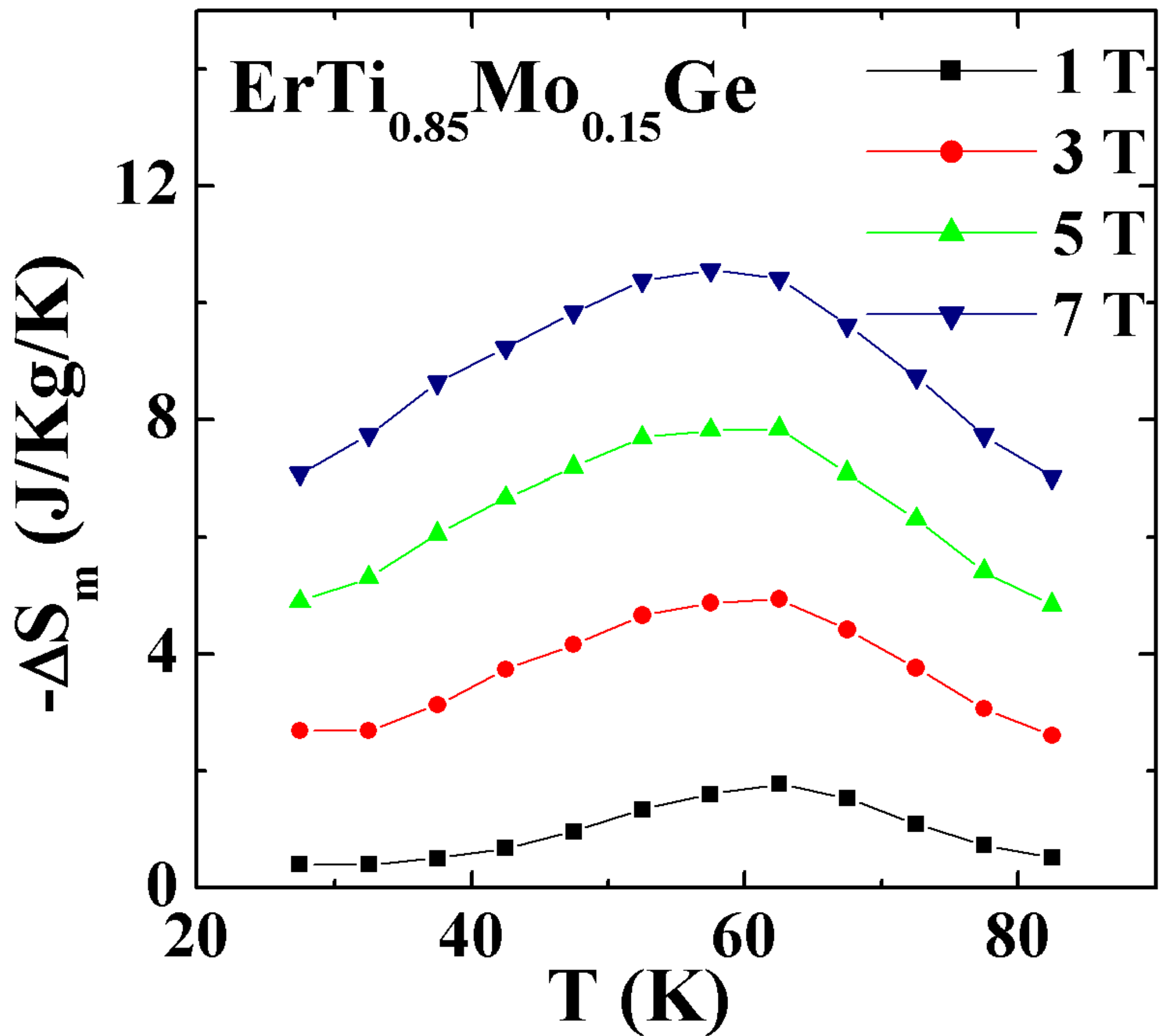

Fig. 11